\documentclass[prd,twocolumn,superscriptaddress]{revtex4-1}
\usepackage{graphicx}

\begin{document}
\title{A perturbative and gauge invariant treatment of gravitational wave memory}

\author{Lydia Bieri}
\email{lbieri@umich.edu}
\affiliation{Dept. of Mathematics, University of Michigan, Ann Arbor, MI 48109-1120, USA}
\author{David Garfinkle}
\email{garfinkl@oakland.edu}
\affiliation{Dept. of Physics, Oakland University, Rochester, MI 48309, USA}
\affiliation{Michigan Center for Theoretical Physics, Randall Laboratory of Physics, University of Michigan, Ann Arbor, MI 48109-1120, USA}

%%%%%%%%%%%%%%%%%%%%%%%%%%%%%%%%%%%%%%%%%%%%%%%%%%%

\date{\today}

%%%%%%%%%%%%%%%%%%%%%%%%%%%%%%%%%%%%%%%%%%%%%%%%%%%
\begin{abstract}
We present a perturbative treatment of gravitational wave memory. 
The coordinate invariance of Einstein's equations leads to a type of gauge invariance in perturbation theory.  As with any gauge invariant theory, results are more clear when expressed in terms of manifestly gauge invariant quantities. 
Therefore we derive all our results from the perturbed Weyl tensor rather than the perturbed metric. 
We derive gravitational wave memory for the Einstein equations coupled to a general energy-momentum tensor that reaches null infinity.

\end{abstract}

%%%%%%%%%%%%%%%%%%%%%%%%%%%%%%%%%%%%%%%%%%%%%%%%%%%%%%%%%%%%%%%%%%%%%%%%%%%%%%

\maketitle

%%%%%%%%%%%%%%%%%%%%%%%%%%%%%%%%%%%%
%%%%%%%%%%%%%%%%%%%%%%%%%%%%%%%%%%%%
\section{Introduction}
\indent

Gravitational wave memory is a distortion in a gravitational wave detector that persists even after the wave has
passed.  As a phenomenon in weak field, slow motion gravity, memory has been well understood since the work of 
Zel'dovich, Polnarev, Braginsky, Grishchuk, and Thorne.\cite{zeldovich,braginsky1,braginsky2}  
This ``linear'' memory is due to a change in the second time derivative of the quadrupole moment 
of the source.  However, Christodoulou\cite{christodoulou} found that there is an additional ``nonlinear'' 
memory due to the energy radiated in gravitational radiation.  The results of \cite{christodoulou}, based on the
global nonlinear stability result of Christodoulou and Klainerman,\cite{globalCK} 
make use of the full nonlinear general theory of 
relativity without any perturbative approximation.  And indeed it is the point of view of \cite{christodoulou} that this
memory is an inherently nonlinear phenomenon that cannot be captured by perturbation theory.  Since the energy of gravitational radiation is second order in gravitational perturbations, it is certainly true that the nonlinear memory of 
\cite{christodoulou} cannot be treated in first order perturbation theory.  Nonetheless, that leaves open the possibility of a treatment in second order perturbation theory.  Furthermore, it has been shown that in the presence of electromagnetic 
fields\cite{zipser,lydia1,lydia2} and neutrinos\cite{lydiaandme} the nonlinear memory depends on the energy radiated to 
infinity by the matter fields in exactly the same way that it depends on the energy radiated in gravitational waves.  In these
cases, the memory due to matter fields is only first order in the gravitational perturbation (though it may be higher order
in the matter fields), so there is the possibility of treating these cases in first order gravitational perturbation theory.
Indeed there have been perturbative treatments of both the memory due to the 
energy of gravitational waves\cite{thorne,will} and
the memory due to neutrinos.\cite{epstein,turner}  One feature of the treatments of \cite{thorne,will,epstein,turner}
is that they use metric perturbations.  However, due to the coordinate invariance of general relativity, metric perturbations
are not gauge invariant quantities.  Thus, any results obtained using metric perturbations must be carefully examined to see whether they are gauge invariant.  Furthermore, as we will show in appendix \ref{metric} there are special difficulties that
arise when using metric perturbations with matter fields whose energy can get out to null infinity.  For that reason, we prefer
an approach using the perturbed Weyl tensor.  Such an approach is gauge invariant from the start and is easily able to handle
stress-energy that gets to null infinity.  Furthermore, when expressed in terms of the electric and magnetic parts of the Weyl tensor there is a close analogy between the first order Einstein equation  and Maxwell's equations.  We exploit this analogy
by using the method of \cite{emanalog} (which treats an electromagnetic analog of gravitational wave memory) to treat
the linearized Einstein equation using the perturbed Weyl tensor.  

From the treatments of \cite{christodoulou} and \cite{jorg} it is clear that gravitational wave memory is a property of 
the asymptotic gravitational field in the limit to null infinity.  We will therefore obtain our results by taking the limit
to null infinity of the linearized Einstein equations for the Weyl tensor.  Our results thus hold for the class of spacetimes where linearized gravity is a good approximation in a neighborhood of null infinity, even if the gravitational field becomes strong in the interior of the spacetime.  In the treatments of \cite{christodoulou} and
\cite{jorg} the results appear to depend on the detailed geometric behavior of a carefully defined optical scalar $u$.  
This seems odd from the perturbative point of view, since in first order perturbation theory the only gauge invariant 
quantity is the perturbed Weyl tensor.  It turns out that our approach yields a geometric system that is essentially
equivalent to those of \cite{christodoulou} and \cite{jorg} but where all quantities are expressed directly in terms
of the perturbed Weyl tensor.

\section{Perturbative treatment}

We work to first order in deviations from Minkowski spacetime.  It will be useful to use both
Cartesian coordinates and spherical coordinates with the following notation: spacetime Cartesian coordinate indices
are denoted by greek letters, spatial Cartesian coordinate indicies by lower case latin letters, and spherical 
coordinate indicies by $r$ if they are in the radial direction and by capital latin letters if they are in the
two-sphere direction. 
The Weyl tensor is given in terms of the Riemann tensor by
\begin{eqnarray}
&{C_{\alpha \beta \gamma \delta}}& = {R_{\alpha \beta \gamma \delta}} 
\nonumber
\\
&-& {\textstyle {1 \over 2}}
\left ( {g_{\alpha \gamma}} {S_{ \delta \beta}} - {g_{\alpha \delta}} {S_{\gamma \beta}} - 
{g_{\beta \gamma}} {S_{\delta \alpha}} + {g_{\beta \delta}} {S_{\gamma \alpha}} \right ) 
\end{eqnarray}
Where $S_{\alpha \beta}$ is defined by
\begin{equation}
{S_{\alpha \beta}} = {R_{\alpha \beta}} - {\textstyle {1 \over 6}} R {g_{\alpha \beta }}
\end{equation}
Since Minkowski spacetime has vanishing Weyl tensor, it follows that in first order perturbation
theory the Weyl tensor is gauge invariant. From the Einstein field equation it follows that
\begin{equation}
{S_{\alpha \beta}} = 8 \pi ( {T_{\alpha \beta}} - {\textstyle {\frac 1 3}} T {g_{\alpha \beta}})
\end{equation}
where $T_{\alpha \beta}$ is the stress-energy tensor.  We will assume that the
stress-energy tensor satisfies the dominant energy condition.  
The Weyl tensor is decomposed into its electric and magnetic parts, which are defined by
\begin{eqnarray}
{E_{ab}} \equiv {C_{atbt}}
\\
{B_{ab}} \equiv {\textstyle {1 \over 2}} {{\epsilon ^{ef}}_a}{C_{efbt}}
\end{eqnarray}
Here $\epsilon _{abc}$ is the spatial volume element and is related to the spacetime volume element by
${\epsilon_{abc}} = {\epsilon_{tabc}}$.  The electric part of the Weyl tensor is important for gravitational
wave memory since for two objects in free fall with a spatial separation $\Delta {x^a}$ we have
\begin{equation}
{\frac {{d^2}\Delta {x^a}} {d{t^2}}} = - {{E^a}_b}\Delta {x^b}
\label{geodev}
\end{equation}
The gravitational wave interferometer is assumed to be at a large distance from the source, and at large 
distances the components of the Weyl tensor fall off at least as fast as $r^{-1}$.  Thus the gravitational wave
memory is essentially the $r^{-1}$ piece of $E_{ab}$ integrated twice with respect to time.  

It will also be useful to decompose the stress-energy tensor into spatial tensors as follows:
\begin{eqnarray}
\mu \equiv {T_{tt}}
\\
{q_a} \equiv {T_{ta}}
\\
{U_{ab}} \equiv {T_{ab}}
\end{eqnarray}

In order to find the behavior of the electric and magnetic parts of the Weyl tensor, we will do the following:
obtain equations of motion (and constraint equations) for these fields, decompose these fields and their equations
into radial quantities and quantities on the two-sphere, and then expand all quantities and equations as power
series in $r^{-1}$.

From the Bianchi identity ${\partial _{[\epsilon }}{R_{\alpha \beta ] \gamma \delta}} =0$
we obtain two constraint equations
\begin{eqnarray}
{\partial ^b}{E_{ab}} = 4 \pi ( {\textstyle {\frac 1 3}}{\partial _a}(2\mu + {{U^c}_c}) - {\partial _t} {q_a})
\label{cnstrE}
\\
{\partial ^b}{B_{ab}} = 4 \pi {{\epsilon ^{ef}}_a} {\partial _e}{q_f}
\label{cnstrB}
\end{eqnarray}
and two equations of motion
\begin{eqnarray}
{\partial _t} {E_{ab}} - {\textstyle {1 \over 2}} {{\epsilon _a}^{cd}}{\partial _c}{B_{db}} 
- {\textstyle {1 \over 2}} {{\epsilon _b}^{cd}}{\partial _c}{B_{da}} 
\nonumber
\\
= 4 \pi 
\left ( {\partial _{(a}}{q_{b)}} - {\textstyle {\frac 1 3}}{\delta _{ab}}{\partial _c}{q^c}
 - {\partial _t} ( {U_{ab}} - {\textstyle {\frac 1 3}}{\delta _{ab}}{{U^c}_c}) \right )
\label{evolveE}
\\
{\partial _t} {B_{ab}} + {\textstyle {1 \over 2}} {{\epsilon _a}^{cd}}{\partial _c}{E_{db}} 
+ {\textstyle {1 \over 2}} {{\epsilon _b}^{cd}}{\partial _c}{E_{da}} 
\nonumber
\\
= 2 \pi 
{{\epsilon _a}^{cd}}{\partial _c}{U_{db}} + 2 \pi 
{{\epsilon _b}^{cd}}{\partial _c}{U_{da}} 
\label{evolveB}
\end{eqnarray}

We now want to decompose the spatial tensors into tensors on the two-sphere.  From the electric part of the Weyl tensor 
$E_{ab}$ we obtain a scalar $E_{rr}$ as well as a vector and a symmetric, trace-free tensor given by
\begin{eqnarray}
{X_A} = {E_{Ar}}
\\
{{\tilde E}_{AB}} = {E_{AB}} - {\textstyle {1 \over 2}} {H_{AB}} {{E_C}^C}
\end{eqnarray}
Here $H_{AB}$ is the metric on the unit two-sphere, and all two-sphere indicies are raised and lowered with this metric.
Similarly, the decomposition of the magnetic part of the Weyl tensor yields $B_{rr}$ and 
\begin{eqnarray}
{Y_A} = {B_{Ar}}
\\
{{\tilde B}_{AB}} = {B_{AB}} - {\textstyle {1 \over 2}} {H_{AB}} {{B_C}^C}
\end{eqnarray}
The decomposition of the spatial vector $q_a$ yields a two-sphere scalar $q_r$ and vector $q_A$, while the decomposition
of the spatial tensor $U_{ab}$ yields two-sphere scalars $U_{rr}$ and $N \equiv {{U^c}_c}$, 
vector ${V_A} \equiv {U_{Ar}}$ and 
a symmetric trace-free tensor 
\begin{equation}
{W_{AB}} = {U_{AB}} - {\textstyle {1 \over 2}} {H_{AB}} {{U_C}^C}
\end{equation} 
Then the constraint equations (eqns. (\ref{cnstrE}) and (\ref{cnstrB})) become
\begin{eqnarray}
{\partial _r}{E_{rr}} + 3 {r^{-1}}{E_{rr}} + {r^{-2}} {D^A}{X_A} 
\nonumber
\\
= 4 \pi
({\textstyle {\frac 1 3}}{\partial _r} (2\mu + N) - {\partial _t}{q_r})
\label{cnstra}
\\
{\partial _r}{B_{rr}} + 3 {r^{-1}}{B_{rr}} + {r^{-2}} {D^A}{Y_A} 
\nonumber
\\
= 4 \pi
{r^{-2}}{\epsilon ^{AB}}{D_A}{q_B}
\label{cnstrb}
\\
{\partial _r}{X_A} + 2 {r^{-1}} {X_A} - {\textstyle {1 \over 2}} {D_A}{E_{rr}} + {r^{-2}} {D^B}{{\tilde E}_{AB}}
\nonumber
\\
= 4 \pi ({\textstyle {\frac 1 3}} {D_A} (2\mu + N) - {\partial _t} {q_A})
\label{cnstrc}
\\
{\partial _r}{Y_A} + 2 {r^{-1}} {Y_A} - {\textstyle {1 \over 2}} {D_A}{B_{rr}} + {r^{-2}} {D^B}{{\tilde B}_{AB}}
\nonumber
\\
= 4 \pi {{\epsilon _A}^B}({D_B} {q_r} - {\partial _r} {q_B})
\label{cnstrd}
\end{eqnarray}
Here $D_A$ is the derivative operator and $\epsilon_{AB}$ is the volume element of the unit two-sphere.  
The evolution equations (eqns. (\ref{evolveE}) and (\ref{evolveB})) become
\begin{eqnarray}
{\partial _t} {B_{rr}} + {r^{-2}} {\epsilon^{AB}}{D_A}{X_B} 
= 4 \pi {r^{-2}} {\epsilon^{AB}}{D_A}{V_B}
\label{evolvea}
\\
{\partial _t} {E_{rr}} - {r^{-2}} {\epsilon^{AB}}{D_A}{Y_B} 
\nonumber
\\
=  4 \pi \left ( {\partial _r}{q_r} - {\partial _t}{U_{rr}} + {\textstyle {\frac 1 3}} {\partial _t} (N-\mu) \right )
\label{eovlveb}
\\
{\partial _t}{Y_A} + {\textstyle {1 \over 2}} {r^{-2}} {\epsilon^{CD}}{D_C}{{\tilde E}_{DA}} 
\nonumber
\\
+ {\textstyle {1 \over 4}} {{\epsilon_A}^C} (3 {D_C}{E_{rr}} - 2 {\partial _r}{X_C})
\nonumber
\\
= 2 \pi {{\epsilon_A}^C} ({\textstyle {\frac 1 2}}{D_C}(3{U_{rr}}-N)-{\partial _r}{V_C})
\nonumber
\\
+ 2 \pi {r^{-2}}{\epsilon ^{BC}}{D_B}{W_{CA}}
\label{evolvec}
\\
{\partial _t}{X_A} - {\textstyle {1 \over 2}} {r^{-2}} {\epsilon^{CD}}{D_C}{{\tilde B}_{DA}} 
\nonumber
\\
- {\textstyle {1 \over 4}}  {{\epsilon_A}^C}(3 {D_C}{B_{rr}} - 2 {\partial _r}{Y_C})
\nonumber
\\
= 2 \pi ({D_A}{q_r}+{\partial _r}{q_A}) - 4 \pi ({r^{-1}}{q_A} + {\partial _t}{V_A})
\label{evolved}
\\
{\partial _t} {{\tilde B}_{AB}} + {\textstyle {1 \over 2}} {{\epsilon_A}^C} 
({D_C}{X_B}+ {r^{-1}}{{\tilde E}_{CB}} - {\partial _r} {{\tilde E}_{CB}}) 
\nonumber
\\
+ {\textstyle {1 \over 2}} {{\epsilon_B}^C} 
({D_C}{X_A} + {r^{-1}}{{\tilde E}_{CA}} - {\partial _r} {{\tilde E}_{CA}}) 
\nonumber
\\
- {\textstyle {1 \over 2}} {H_{AB}} {\epsilon^{CD}}
{D_C}{X_D}
\nonumber
\\
= 2 \pi {{\epsilon_A}^C}({D_C}{V_B} + {r^{-1}}{W_{CB}}- {\partial _r}{W_{CB}})
\nonumber
\\
+ 2 \pi {{\epsilon_B}^C}({D_C}{V_A} + {r^{-1}}{W_{CA}}- {\partial _r}{W_{CA}})
\nonumber
\\
+ 2 \pi {H_{AB}}{\epsilon^{CD}}{D_C}{V_D}
\label{evolvee}
\\
{\partial _t} {{\tilde E}_{AB}} - {\textstyle {1 \over 2}} {{\epsilon_A}^C} 
({D_C}{Y_B}- {\partial _r} {{\tilde B}_{CB}}+{r^{-1}} {{\tilde B}_{CB}}) 
\nonumber
\\
- {\textstyle {1 \over 2}} {{\epsilon_B}^C}  
({D_C}{Y_A}- {\partial _r} {{\tilde B}_{CA}}+{r^{-1}} {{\tilde B}_{CA}}) 
\nonumber
\\
- {\textstyle {1 \over 2}} {H_{AB}}{\epsilon^{CD}}{D_C}{Y_D}
\nonumber
\\
= 4 \pi ( {D_{(A}}{q_{B)}} - {\partial _t}{W_{AB}} - {\textstyle {1 \over 2}} 
{H_{AB}} {D_C}{q^C} )
\label{evolvef}
\end{eqnarray}

The next step is to consider the behavior of the fields near null infinity.  In appendix \ref{falloff} we show that 
the stress-energy behaves as follows:
\begin{equation}
\mu = -{q_r} = {U_{rr}} = N =  L {r^{-2}} + \dots
\label{nullstress}
\end{equation}
with all other components of the stress-energy falling off more rapidly.
Here $\dots$ means ``terms higher order in $r^{-1}$'' and $L$ is a function of $u$ and the
two-sphere coordinates.  In physical terms, the quantity $L$ is the power radiated per unit solid angle.  
In appendix \ref{falloff} we show that the electric and magnetic parts of the Weyl tensor behave as follows: 
\begin{eqnarray}
{{\tilde E}_{AB}} = {e_{AB}}r + \dots
\\
{{\tilde B}_{AB}} = {b_{AB}}r + \dots
\\
{X_A} = {x_A}{r^{-1}} + \dots
\\
{Y_A} = {y_A}{r^{-1}} + \dots
\\
{E_{rr}} = P {r^{-3}} + \dots
\\
{B_{rr}} = Q {r^{-3}} + \dots
\end{eqnarray}
Here the coefficient tensor fields are functions of $u$ and the two-sphere coordinates.  
Furthermore, in the limit as $|u| \to \infty$ the only one of these coefficient tensor fields that 
does not vanish is $P$. Note that because
of the relation between Cartesian and spherical coordinates ${\tilde E}_{AB}$ behaving like $r$ corresponds
to Cartesian components of the electric part of the Weyl tensor behaving like $r^{-1}$.  

Now keeping only the dominant terms in eqns. (\ref{cnstra}-\ref{cnstrd}) we obtain
\begin{eqnarray}
- {\dot P} + {D^A}{x_A} = - 8 \pi L
\label{scricna}
\\
- {\dot Q} + {D^A}{y_A} = 0
\label{scricnb}
\\
- {{\dot x}_A} + {D^B}{e_{AB}} = 0
\label{scricnc}
\\
- {{\dot y}_A} + {D^B}{b_{AB}} = 0
\label{scricnd}
\end{eqnarray}
Here an overdot means derivative with respect to $u$.  Similarly, 
keeping only the dominant terms in eqns. (\ref{evolvea}-\ref{evolvef}) yields
\begin{eqnarray}
{\dot Q} + {\epsilon^{AB}}{D_A}{x_B} = 0 
\label{scrieva}
\\
{\dot P} - {\epsilon^{AB}}{D_A}{y_B} = 8 \pi L 
\label{scrievb}
\\
{{\dot y}_A} + {\textstyle {1 \over 2}} {\epsilon^{CD}}{D_C}{e_{DA}} +  {\textstyle {1 \over 2}} 
{{\epsilon_A}^C}{{\dot x}_C} = 0
\label{scrievc}
\\
{{\dot x}_A} - {\textstyle {1 \over 2}}{\epsilon^{CD}} {D_C}{b_{DA}} -  {\textstyle {1 \over 2}} 
{{\epsilon_A}^C}{{\dot y}_C} = 0
\label{scrievd}
\\
{{\dot b}_{AB}} + {{\epsilon_A}^C}{{\dot e}_{CB}} = 0
\label{scrieve}
\\
{{\dot e}_{AB}} - {{\epsilon_A}^C}{{\dot b}_{CB}} = 0
\label{scrievf}
\end{eqnarray}

Note that eqn. (\ref{scrievf}) is redundant, since it is equivalent to eqn. (\ref{scrieve}).  
Since $e_{AB}$ and $b_{AB}$ vanish as
$u \to - \infty$, it follows from eqn. (\ref{scrieve}) that
${b_{AB}} = - {{\epsilon_A}^C}{e_{CB}}$.  This can be used to 
eliminate $b_{AB}$ from eqns. (\ref{scricnd}) and (\ref{scrievd}) which then become
\begin{eqnarray}
{{\dot y}_A} +  {\epsilon^{CD}}{D_C}{e_{DA}} = 0
\label{reduce1a}
\\
{{\dot x}_A} - {\textstyle {1 \over 2}} {D^C}{e_{CA}} -  {\textstyle {1 \over 2}} 
{{\epsilon_A}^C}{{\dot y}_C} = 0
\label{reduce1b}
\end{eqnarray}
Combining eqn. (\ref{reduce1a}) with eqn. (\ref{scrievc}) then yields
\begin{equation}
{{\dot y}_A} + {{\epsilon_A}^B}{{\dot x}_B} = 0
\label{reduce2}
\end{equation}
However, since $x_A$ and $y_A$ vanish as $u \to - \infty$.  It then follows from eqn. (\ref{reduce2}) that  
\begin{equation}
{y_A} = - {{\epsilon_A}^B}{x_B}
\end{equation}
Thus, we can eliminate $y_A$ from eqns. (\ref{scricnb}) and (\ref{scrievb}) which then become
\begin{eqnarray}
{\dot Q} + {\epsilon^{AB}}{D_A}{x_B} = 0 
\\
{\dot P} - {D^A}{x_A} =  8 \pi L
\end{eqnarray}
But these equations are then redundant, since they are equivalent to 
eqns. (\ref{scrieva}) and (\ref{scricna}) respectively.  Thus the only 
independent quantities are ${e_{AB}}, \, {x_A}, \, P, \, Q$ and $L$.  These quantities satisfy the following equations
\begin{eqnarray}
{D^B}{e_{AB}} = {{\dot x}_A}
\label{reduce3a}
\\
{\epsilon^{BC}}{D_B}{e_{CA}} = {{\epsilon_A}^C}{{\dot x}_C}
\label{reduce3b}
\\
{D_A}{x^A} = {\dot P} - 8 \pi L
\label{reduce3c}
\\
{\epsilon^{AB}}{D_A}{x_B} = - {\dot Q}
\label{reduce3d}
\end{eqnarray}

Now let's consider how to use eqns. (\ref{reduce3a}-\ref{reduce3d}) to find the memory.  
Recall that $e_{AB}$ is (up to a factor involving the distance
and the initial separation) the second time derivative of the separation of the masses.  Thus we want to integrate $e_{AB}$
twice with respect to $u$.  Define the velocity tensor $v_{AB}$, memory tensor $m_{AB}$ and a tensor $z_A$ by
\begin{eqnarray}
{v_{AB}} \equiv {\int _{- \infty} ^u} {e_{AB}} du
\\
{m_{AB}} \equiv {\int _{- \infty} ^\infty} {v_{AB}} du
\\
{z_A} \equiv {\int _{- \infty} ^\infty} {x_A} du
\end{eqnarray} 
Now consider two masses in free fall whose initial separation is $d$ in the $B$ direction.  Then after the wave has passed
they will have an additional separation.  Call the component of that additional separation in the $A$ direction $\Delta d$.
Then it follows from eqn. (\ref{geodev}) that
\begin{equation} \label{mem*2}
\Delta d = - {\frac d r} {{m^A}_B}
\end{equation}
To find $m_{AB}$ we first integrate eqns. (\ref{reduce3a}) and (\ref{reduce3b}) to obtain
\begin{eqnarray}
{D^B}{v_{AB}} = {x_A}
\\
{\epsilon^{BC}}{D_B}{v_{CA}} = {{\epsilon_A}^C}{x_C}
\end{eqnarray}
Then integrating again from $-\infty$ to $\infty$ we obtain
\begin{eqnarray}
{D^B}{m_{AB}} = {z_A}
\label{system1a}
\\
{\epsilon^{BC}}{D_B}{m_{CA}} = {{\epsilon_A}^C}{z_C}
\label{system1b}
\end{eqnarray}
Now integrating eqns. (\ref{reduce3c}) and (\ref{reduce3d}) from $-\infty$ to $\infty$ yields
\begin{eqnarray}
{D_A}{z^A} = \Delta P - 8 \pi F
\label{system1c}
\\
{\epsilon^{AB}}{D_A}{z_B} = 0
\label{system1d}
\end{eqnarray}
where the quantities $\Delta P$ and $F$ are defined by
$\Delta P = P(\infty) - P(-\infty)$ and 
$F = {\int _{-\infty} ^\infty } L du$.  In physical terms, $F$ is the amount of energy radiated per unit solid angle.
In deriving eqn. (\ref{system1d}) we have used the fact that $Q$ vanishes in the limit as $|u| \to \infty$.  Since $z_A$ is curl-free, there must be a scalar $\Phi$ such that 
${z_A} = {D_A} \Phi$.  Then using eqns. (\ref{system1c}) and (\ref{system1a}) we find
\begin{eqnarray}
{D_A}{D^A}\Phi =  \Delta P - 8 \pi  F 
\label{system2a}
\\
{D^B}{m_{AB}} = {D_A}\Phi
\label{system2b}
\end{eqnarray}
By expanding in spherical harmonics, one can show that the consistency of eqns. (\ref{system2a}-\ref{system2b}) requires that
the right hand side of eqn. (\ref{system2a}) has vanishing $\ell =0 $ piece and vanishing $\ell =1$ piece.  For any 
quantity on the 2-sphere, we will adopt the notation that a subscript $[1]$ means the $\ell =0$ and $\ell =1$ part of
that quantity.  It follows from eqn. (\ref{system2a}) that $\Phi$ consists of two pieces $\Phi = {\Phi_1} + {\Phi_2}$
satisfying the following equations:
\begin{eqnarray}
{D_A}{D^A}{\Phi_1} =   \Delta P - {{(\Delta P)}_{[1]}} 
\\
{D_A}{D^A}{\Phi_2} = - 8 \pi (F - {F_{[1]}})  
\end{eqnarray}
and that ${m_{AB}}={m_{1AB}}+{m_{2AB}}$ with ${D^B}{m_{1AB}} = {D_A}{\Phi_1}$ and correspondingly for $m_{2AB}$.  
In \cite{christodoulou} $m_{1AB}$ is called the ordinary memory and $m_{2AB}$ is called the nonlinear memory.

We now work out explicitly the dependence of $m_{2AB}$ on $F$.   
Eqns. (\ref{system2a}-\ref{system2b}) are equivalent to eqn. (10-12) of \cite{christodoulou}.  The solution is thus 
the one given in that paper.  Nonetheless, we will find it helpful to derive a different formula for $m_{2AB}$ using
an expansion in spherical harmonics.  (Note that the same method yields the dependence of the ordinary memory $m_{1AB}$ 
on $\Delta P$).  For an explicit comparison with the formula of \cite{christodoulou} we also provide a
Green's function method in Appendix \ref{Green}.
We have
\begin{eqnarray}
- 8 \pi ( F - {F_{[1]}}) = {\sum _{\ell >1}} {a_{\ell m}} {Y_{\ell m}}
\label{ylmsource}
\\
{m_{2AB}} = {\sum _{\ell >1}} {b_{\ell m}} ({D_A}{D_B}{Y_{\ell m}} - {\textstyle {1\over 2}} {H_{AB}}  {D_C}{D^C}{Y_{\ell m}} )
\label{ylmmemory}
\end{eqnarray}
Then using eqns. (\ref{system2a}) and (\ref{system2b}) we find the the expansion coefficients $b_{\ell m}$ are given by
\begin{equation}
{b_{\ell m}} = {{2 {a_{\ell m}}} \over {(\ell -1)\ell (\ell +1)(\ell +2)}}
\end{equation}
Thus the result is that the memory tensor is given by the expression in 
eqn. (\ref{ylmmemory}) where the expansion coefficients are given 
in terms of the source by 
\begin{equation}
{b_{\ell m}} = {{- 2 } \over {(\ell -1)\ell (\ell +1)(\ell +2)}}
\int d \Omega {Y^* _{\ell m}} 
8 \pi F
\label{ylmcoeff}
\end{equation}
Note that $b_{\ell m}$ is defined only for $\ell \ge 2$ and that eqn. (\ref{ylmcoeff}) has a large power of $\ell$ in
the denominator.  It then seems likely that to a very good approximation gravitational wave memory is given by its
$\ell=2$ part.  We therefore expect that the size of the gravitational wave memory effect is essentially determined
by the amount of energy radiated by the source in the $\ell = 2$ modes.  

\section{Conclusions}

Though the most rigorous treatments of gravitational wave memory use the full nonlinear Einstein field equations, we
have shown that many of the interesting properties of gravitational wave memory can be captured in first order perturbation
theory.  Since all results are expressed in terms of gauge invariant quantities, the physical nature of gravitational wave 
memory is made more clear in our treatment than in those treatments that rely on metric perturbations.  In particular, we have 
shown that there are indeed two types of gravitational wave memory, but that rather than calling them ``linear'' and 
``nonlinear'' memory, it is perhaps more clear to call them ``ordinary'' and ``null.''  The ordinary memory has to do with
changes in the $E_{rr}$ component of the Weyl tensor between initial and final states of the system.  For slowly moving sources,
this ordinary memory can be expressed in terms of the difference between the second time derivative of the source quadrupole 
moment between the initial and final states.  The null memory is due to the energy-momentum tensor of fields that get out to null infinity, and can be expressed in terms of the energy radiated per unit solid angle.  These two types of memory are 
completely analogous to our results\cite{emanalog} for the ``memory'' of test charges propelled by an electromagnetic wave.  

We have performed an expansion in spherical harmonics of the memory effect and shown that it is predominantly a quadrupolar 
(that is $\ell =2$) effect.  Thus the null memory effect is mainly due to the part of the radiated energy that is in 
the $\ell=2$ modes.  This provides a simple method for using the results of numerical simulations of core collapse supernovae 
and of gamma ray bursts to obtain estimates of the size of the gravitational wave memory effect from each of these systems.

Our perturbative results are in complete agreement with the fully nonlinear treatments of gravitational wave memory due to 
the energy of electromagnetic fields or neutrinos.  However, we do not yet provide a perturbative derivation of the memory due to the energy radiated in gravitational waves.  This is because our treatment is in first order perturbation theory,
while the effects of \cite{christodoulou} do not appear until second order in perturbation theory.  The case treated in
\cite{christodoulou} is thus an example where first order perturbation theory is not adequate, even in a neighborhood of null infinity.  We expect that our methods can be generalized to second order perturbation theory; however such a generalization is not completely straightforward because the issue of gauge invariance is more complicated in second order perturbation theory since the perturbed Weyl tensor is no longer gauge invariant.  One possible approach would be to treat the second order perturbations as sourced by an effective gravitational stress-energy that is quadratic in the first order perturbations. It would be interesting to perform such a second order analysis and compare to the fully nonlinear results of \cite{christodoulou}.

\section*{Acknowledgements}
DG was supported by NSF grants PHY-0855532 and PHY-1205202 to Oakland University. 
LB was supported by NSF grants DMS-0904760 and DMS-1253149 to The University of Michigan.
This material is based upon work supported by the National Science Foundation under Grant No.
0932078000, while the authors were in residence at the Mathematical Sciences Research Institute in Berkeley,
California during the Fall Semester of 2013.
We would like to thank Bob Wald for helpful discussions.
 
\appendix

\section{Difficulties with metric perturbations}
\label{metric}

We now consider how fields whose stress-energy can get to null infinity create difficulties with the usual metric perturbation formalism.  Recall that the metric is written to first order as
$ {g_{\mu \nu}} = {\eta _{\mu \nu}} + {h_{\mu \nu}} $ where ${\eta _{\mu \nu}}$ is a flat metric and 
$h_{\mu \nu}$ is small.  The coordinate invariance of general relativity leads to a gauge invariance under ${h_{\mu \nu}} \to {h_{\mu \nu}} + 2 {\partial _{(\mu}}{\xi _{\nu )}}$ for any vector field $\xi _\nu$.  The usual Lorentz gauge 
condition is ${\partial ^\mu}{{\bar h}_{\mu \nu}}=0$ where the trace reversed metric perturbation is
defined by ${{\bar h}_{\mu \nu}} = {h_{\mu \nu}} - {\frac 1 2} h {\eta _{\mu \nu}}$.
Then the first order Einstein field equations become
\begin{equation}
{\partial ^\alpha}{\partial _\alpha}{{\bar h}_{\mu \nu}} = - 16 \pi {T_{\mu \nu}}
\label{lineinstein}
\end{equation}
The retarded solution of eqn. (\ref{lineinstein}) is
\begin{equation}
{{\bar h}_{\mu \nu}}(t,{\vec r}) = 4 \int {\frac {{d^3}{R}\;  {T_{\mu \nu}}(T,{\vec R})} 
{| {\vec r} -{ \vec R}|}}
\label{ret}
\end{equation}

Here the point $(T,{\vec R})$ is on the past light cone of the point $(t,{\vec r})$ and therefore we
have 
\begin{equation}
t - T = |{\vec r}-{\vec R}|
\end{equation}
We would like to know the behavior of $h_{\mu \nu}$ near null infinity, that is for large $r$ and
finite $u$.  It  is natural to assume that $R \ll r$ which leads to the approximation
\begin{eqnarray}
{\frac 1 { |{\vec r}-{\vec R}|}} = {\frac 1 r}
\label{largera}
\\
T = u + {\hat r}\cdot {\vec R}
\label{largerb}
\end{eqnarray}
However, it turns out that things are not quite so simple.  First consider the case where the stress energy is that of a timelike or null particle traveling along a geodesic.  For a timelike geodesic, we do have $R \ll r$.  And for a null geodesic we have $R \ll r$ for all values of $\hat R$ 
{\it except} ${\hat R} = {\hat r}$.  Thus, in order to use the approximation of eqns. (\ref{largera}-\ref{largerb}) for stress-energy that can get to null infinity, we have two choices: either only calculate the field at points of null infinity that are not approached by the stress-energy, as is done in \cite{wald}, or define the metric perturbation using a limiting procedure as follows:  let $\theta$ be the angle between $\hat r$ and $\hat R$.  Then in the integral over all 
$\vec R$ we exclude the region where $\theta < \epsilon$ and then take the limit as $\epsilon \to 0$.   Thus we are tentatively led to the following expression for the metric perturbation near null infinity
\begin{equation}
{{\bar h}_{\mu \nu}} = {\frac 4 r} {\lim _{\epsilon \to 0}} {\int _{\theta > \epsilon}} {d^3} R \; 
{T_{\mu \nu}}(u + {\hat r}\cdot {\vec R},{\vec R})
\label{retmet}
\end{equation} 
However, this expression only works if the limit exists, and we will now argue that it does not.  Consider null geodesics emitted near the center that cross the null plane given by eqn. (\ref{largerb}) at a small value of $\theta$.  Then for a given range of times of emission, the value of $R$ at which the geodesics intersect the null plane, and the range $dR$ go like $\theta ^{-2}$.  (This is in contrast to the behavior of timelike geodesics, where $R$ has a limiting value at small $\theta$).  Now we can write ${d^3}R$ as ${R^2} d R d \Omega$ where $d\Omega$ is the measure on the unit two-sphere.   Since at large $R$ the stress-energy goes like $R^{-2}$ we find that ${T_{\mu \nu}}{R^2} dR $ goes like $\theta ^{-2}$.  Since for small $\theta$ we have that 
$d \Omega = 2 \pi \theta d \theta$, it follows that the quantity  in eqn. (\ref{retmet})  whose limit we are trying to take goes as $\ln \epsilon$ and therefore the limit as $\epsilon \to 0$ does not exist.   

Note that this argument is quite general, as it depends only on the properties of the wave equation and a source that can get to null infinity.  Thus, for example, it also applies to Maxwell's equations using a vector potential in Lorentz gauge.  If one uses a charge current that can get to null infinity, then the vector potential in Lorentz gauge is not well behaved.  One might worry that the argument is too general for the following reason: from Maxwell's equations it follows that the Cartesian components of the electric and magnetic fields satisfy the wave equation with source.  Thus, it seems that we might be led to believe that the electric and magnetic fields are not well behaved.  However, in this case one can show that the source vanishes at $\theta =0$ and therefore that the limit as $\epsilon \to 0$ in the analog of eqn. (\ref{retmet}) exists.  Similar considerations apply to the sources for the electric and magnetic parts of the Weyl tensor and the wave equations that they satisfy.  

Since the standard gauge for linearized gravity does not work when the sources can get to null infinity, one has a choice of either finding a better gauge or working with gauge invariant quantities.   We have chosen the second approach, though there is certainly the possibility that the first approach might also work.  Since the usual metric perturbation approach is problematic when treating gravitational wave memory with sources that can get to null infinity, what attitude should one take towards such treatments of memory as those of \cite{epstein,turner,thorne,will} that use this approach?  First note that the results of these references are expressed in terms of the transverse traceless part of the metric perturbation.  It is certainly possible that the transverse traceless part of the metric perturbation is better behaved than the metric perturbation itself.  That is, in this case taking the transverse traceless part may also amount to discarding those parts of the metric perturbation that are ill behaved.  However, the general approach of metric perturbation theory implicitly assumes that  one is using a gauge in which the metric perturbations are well behaved.   When this is not the case, we prefer a gauge invariant method.

\section{Behavior of the fields near null infinity}
\label{falloff}

We would like to know how the stress-energy and the Weyl tensor behave near null infinity, that is for large $r$ and finite 
retarded time $u=t-r$.  We define the advanced time $v=t+r$ and the future directed null vectors 
${\ell _\mu} = - {\partial _\mu} u $ and ${n _\mu} = - {\partial _\mu} v$.  We will assume that the Cartesian components of
both the stress-energy and the Weyl tensor can be expanded near null infinity in power series in $r^{-1}$ with coefficients
that are smooth functions of $u$ and the angular coordinates.  We will also assume that the stress-energy satisfies the dominant energy condition.  Since the power radiated per unit solid angle is the limit as $r \to \infty$ of
$- {r^2} {T_{tr}}$ and since we want that limit to exist and be non-vanishing, we will assume that the stress-energy falls off like $r^{-2}$.  Furthermore, since only a finite amount of mass can be radiated, we will assume that this $r^{-2}$ piece of the stress-energy goes to zero at large $|u|$.  It then 
follows from the properties of the angular coordinates that 
\begin{equation}
{\partial _\alpha}{T_{\mu \nu}} = - {\ell _\alpha} {\frac \partial {\partial u}} {T_{\mu \nu}} \; + \; O({r^{-3}}) \; \; \; .
\label{gradstress}
\end{equation}
But the stress-energy is conserved, and its $r^{-2}$ piece cannot have any part that is unchanging in $u$, so it follows that 
${\ell ^\mu} {T_{\mu \nu}}$ must be of order $r^{-3}$.  Now using a basis that consists of $\ell _\mu , \, {n_\mu}$ and 
two unit vectors in the angular directions, we find that the stress-energy must take the form
\begin{equation}
{T_{\mu \nu}} = A {\ell _\mu} {\ell _\nu} + 2 {\ell _{(\mu}}{B_{\nu )}} + {C_{\mu \nu}} \; + \; O({r^{-3}})
\end{equation}
where $B_\mu$ and $C_{\mu \nu}$ have components only in the angular directions.  However, the dominant energy condition implies that the stress-energy tensor contracted with any timelike or null vector must yield a timelike or null vector, from which it follows that $B_\mu$ and $C_{\mu \nu}$ must vanish.  Thus, we find that the stress-energy takes the form
\begin{equation}
{T_{\mu \nu}} = {r^{-2}} L {\ell _\mu} {\ell _\nu}  \; + \; O({r^{-3}})
\end{equation}  
from which eqn. (\ref{nullstress}) follows.

Now, we consider the behavior of the Weyl tensor near null infinity.  Contracting the Bianchi identity
${\partial _{[\epsilon }}{R_{\alpha \beta ] \gamma \delta}} =0$ we obtain
\begin{equation}
{\partial ^\epsilon }{R_{\epsilon \alpha \beta \gamma}} = {\partial _\beta}{R_{\gamma \alpha}} - 
{\partial _\gamma}{R_{\beta \alpha}} \; \; \; .
\label{cBianchi}
\end{equation} 
Now acting on the Bianchi identity with $\partial ^\epsilon$ and using eqn. (\ref{cBianchi}) we obtain
\begin{eqnarray}
{\partial ^\epsilon}{\partial _\epsilon}{R_{\alpha \beta \gamma \delta}} = 
{\partial _\alpha}{\partial _\gamma}{R_{\delta \beta}} 
\; - \; {\partial _\alpha}{\partial _\delta}{R_{\gamma \beta}}
\nonumber
\\
\; - \; {\partial _\beta}{\partial _\gamma}{R_{\delta \alpha}}
\; + \; {\partial _\beta}{\partial _\delta}{R_{\gamma \alpha}} \; \; \; .
\label{waveRiemann}
\end{eqnarray}
Thus the Riemann tensor satisfies the wave equation with a source that involves second derivatives of the 
stress-energy tensor.  It then follows from the treatment of appendix \ref{metric} that the Riemann tensor falls off
like $r^{-1}$ at null infinity.  Since the stress-energy falls off like $r^{-2}$ it then follows that the Weyl tensor, and
therefore its electric and magnetic parts, falls off like $r^{-1}$.  We now consider the consequences of the assumption
that the electric and magnetic parts of the Weyl tensor can be expressed as power series in $r^{-1}$ with coefficients that
are smooth functions of $u$ and the angular coordinates.  First note that the spatial derivative of $u$ is 
${\partial _a} u = - {r_a}$ where $r_a$ is the unit spatial vector in the outgoing radial direction.  It then follows from
the same reasoning that led to eqn. (\ref{gradstress}) that 
\begin{equation}
{\partial _c}{E_{ab}} = - {r_c} {\frac \partial {\partial u}} {E_{ab}} \; + \; O({r^{-2}})
\label{gradWeyl}
\end{equation}
But the electric part of the Weyl tensor satisfies
eqn. (\ref{cnstrE}), from which it follows using eqn.(\ref{gradWeyl}) that 
$E_{ra}$ is of order $r^{-2}$.  Now define ${v_a} \equiv {E_{ra}}$.  Then we have  
\begin{equation}
{\partial _c}{v_a} = - {r_c} {\frac \partial {\partial u}} {v_a} \; + \; O({r^{-3}})
\label{gradWeyl2}
\end{equation}
But contracting eqn. (\ref{cnstrE}) with $r^a$ we obtain
\begin{equation}
{\partial ^a}{v_a} = - {r^{-1}}{v_r} + 4 \pi \left ( {\textstyle {\frac 1 3}}{\partial _r}(2 \mu + N) - {\partial _t}
{q_r} \right )
\label{divEr}
\end{equation}
However, the right hand side of eqn. (\ref{divEr}) is $O({r^{-3}})$ and it therefore follows from eqn. (\ref{gradWeyl2}) 
that $E_{rr}$ is $O({r^{-3}})$.  The same reasoning applies to $B_{ab}$ using eqn. (\ref{cnstrB}).  
So we find that $B_{ar}$ is $O({r^{-2}})$ and $B_{rr}$ is $O({r^{-3}})$.

Now we consider the behavior of the Weyl tensor at large $|u|$.  We will assume that at both early and late times the 
matter consists of widely separated objects moving at constant velocity.  Therefore the Weyl tensor is a linear combination
of translated and boosted linearized Schwarzschild perturbations.  Note that in its rest frame the Weyl tensor of 
Schwarzschild falls off as $r^{-3}$.  This property also holds under translations and boosts.  It then immediately follows
that at large $|u|$ the quantities ${e_{AB}}, \, {b_{AB}}, \, {x_A}$ and $y_A$ all vanish since these quantities correspond
to pieces of the Weyl tensor that fall off as $r^{-1}$ and $r^{-2}$.  In its rest frame, the Weyl tensor of Schwarzschild is
purely electric.  The boost does introduce nonzero components of the magnetic part of the Weyl tensor.  However, the $B_{rr}$
component remains zero. It then follows that in the limit as $|u| \to \infty$ the quantity $Q$ vanishes.  
Thus of all the asymptotic Weyl tensor
fields that we use, the only one that does {\it not} vanish in the limit as $|u| \to \infty$ is $P$.

\section{Green's function for memory}
\label{Green}

We want a solution of the system
\begin{eqnarray}
{D^B}{m_{2AB}}={D_A}{\Phi_2} 
\label{systema}
\\
{D^A}{D_A}{\Phi_2}=8\pi ({F_{[1]}}-F)
\label{systemb}
\end{eqnarray}
which we will find using a Green's function method.  
Note that the system is unchanged by adding a constant to $\Phi_2$ so we will specify $\Phi_2$ by imposing the
condition that its average value vanishes.
First use the Ansatz
\begin{equation}
{m_{2AB}}={D_A}{D_B}J - {\textstyle {\frac 1 2}} {H_{AB}}{D_C}{D^C} J
\label{ansatzm}
\end{equation}
Note that the right hand side of eqn. (\ref{ansatzm}) vanishes for $J$ any combination of $\ell=0$ and
$\ell =1$ spherical harmonics.  We will therefore specify $J$ by also imposing the condition that
${J_{[1]}}=0$.  
Then applying eqn. (\ref{systema}) to eqn. (\ref{ansatzm}) we obtain
\begin{equation}
{D_A}({D_B}{D^B}J + 2 J) = 2 {D_A}{\Phi_2}
\end{equation}
which then using the conditions that both $J_{[1]}$ and the average value of $\Phi_2$ vanish yields
\begin{equation}
{D_B}{D^B}J + 2 J = 2 {\Phi_2}
\end{equation}
Now suppose that we want the value of $J$ at a point $p$ on the two-sphere.  We introduce the usual $(\theta,\phi)$ coordinate
system on the two-sphere, with the point $p$ at $\theta=0$.  We also introduce the quantities $x$ and $C$ given by
$x=\cos \theta$ and
\begin{equation}
C = (1-x) \ln (1-x)
\end{equation}
Then we have
\begin{equation}
{D_A}{D^A} C = - 2 C + S
\end{equation}
where the quantity $S$ is given by
\begin{equation}
S=1 + 3 x + 2 \ln(1-x)
\end{equation}
and satisfies
\begin{equation}
{D^A}{D_A} S = -2(1+3x)
\end{equation}
We then have
\begin{eqnarray}
0 = {\lim _{\epsilon \to 0}} {\int _{\theta >\epsilon}} {D_A} ( {\Phi_2}{D^A}C - C{D^A}{\Phi_2}) 
\nonumber
\\
= {\lim _{\epsilon \to 0}} {\int _{\theta >\epsilon}} ({\Phi_2}{D_A}{D^A} C - C {D_A}{D^A} {\Phi_2})
\nonumber
\\
= {\lim _{\epsilon \to 0}} {\int _{\theta >\epsilon}} ({\textstyle {\frac 1 2}}{D_A}{D^A}J + J)(-2C + S) 
\nonumber
\\
+ \int C 8 \pi (F-{F_{[1]}})
\end{eqnarray}
Here the integral sign with no subscript denotes an integral over the two-sphere, while with the subscript $\theta >\epsilon$
the integral is done over only that part of the two-sphere where $\theta >\epsilon$.  In both cases the integral is done
with the usual two sphere volume element.  
However, we also have
\begin{eqnarray}
&{\lim _{\epsilon \to 0}} {\int _{\theta >\epsilon}}& ({\textstyle {\frac 1 2}}{D_A}{D^A}J + J)(-2C + S)
\nonumber
\\
&=& {\lim _{\epsilon \to 0}} {\int _{\theta >\epsilon}} {D_A}(({\textstyle {\frac 1 2}}S-C ){D^A}J 
- J {D^A}({\textstyle {\frac 1 2}}S-C ) )
\nonumber
\\
&+& {\lim _{\epsilon \to 0}} {\int _{\theta >\epsilon}}   J({D_A}{D^A}({\textstyle {\frac 1 2}}S-C ) -2C + S)
\nonumber
\\
&=& {\lim _{x \to 1}} 2 \pi (1-{x^2}) \left [ ({\textstyle {\frac 1 2}} S - C ) {\frac {\partial J} {\partial x}}
- J {\frac \partial {\partial x}} ({\textstyle {\frac 1 2}} S - C ) \right ]
\nonumber 
\\
&+& {\lim _{\epsilon \to 0}} {\int _{\theta >\epsilon}}   J (-(1+3x))
\nonumber 
\\
&=& 4 \pi J (p)
\end{eqnarray}
We therefore obtain
\begin{equation}
J(p) = - 2 \int C  (F-{F_{[1]}})
\label{Jp}
\end{equation}
Note that any point on the two-sphere can be represented as a unit vector in Euclidean 3-space.  Then letting 
$\xi$ be the point $p$ and $\xi '$ be the point that we are integrating over, we have
$x=<\xi, {\xi '}>$ where $<,>$ denotes the Euclidean inner product.  Thus, eqn. (\ref{Jp}) can be written as
\begin{equation}
J(\xi) = - 2 \int (F -{F_{[1]}})({\xi '}) (1 - <\xi, {\xi '}>) \ln (1 - <\xi, {\xi '}>)
\end{equation}
It then follows from eqn. (\ref{ansatzm}) that for any vectors $v^A$ and $w^A$ we have
\begin{eqnarray}
{v^A}{w^B}{m_{2AB}} =  &-& 2 \int (F -{F_{[1]}})({\xi '}) \biggl [ 
{\frac {<v,{\xi '}><w,{\xi '}>} {1 - <\xi, {\xi '}>}}
\nonumber 
\\
 &-& {\textstyle {\frac 1 2}} <v,w>(1+<\xi, {\xi '}>) \biggr ]
\end{eqnarray}

\end{document}